\documentclass[12pt]{iopart}

\usepackage[utf8]{inputenc}
\usepackage[UKenglish]{babel}
\usepackage{graphicx}
\usepackage{braket}
\usepackage{cite}

\usepackage{xcolor}

\begin{document}

\title[Josephson junction cavity systems as cousins of the quantum optical micromaser]{Josephson junction cavity systems as cousins of the quantum optical micromaser}

\author{Simon Dambach$^1$, Andrew D Armour$^2$, Björn Kubala$^1$ and Joachim Ankerhold$^1$}

\address{$^1$ Institute for Complex Quantum Systems and Center for Integrated Quantum Science and Technology, Ulm University, Albert-Einstein-Allee 11, 89081 Ulm, Germany}
\address{$^2$ Centre for the Mathematics and Theoretical Physics of Quantum Non-Equilibrium Systems and School of Physics and Astronomy, University of Nottingham,
Nottingham NG7 2RD, United Kingdom}

\ead{simon.dambach@uni-ulm.de}

\vspace{10pt}

\begin{indented}
\item[]8th August 2019
\end{indented}

\begin{abstract}
We explore the similarities and differences between simple theoretical models developed to describe the quantum optical micromaser and Josephson-photonics devices. Whilst the micromaser has long been recognised as an important model for the dynamics of open quantum systems far from equilibrium, so-called Josephson-photonics devices are a recently developed form of superconducting quantum circuit in which the quantum transport of charges through a voltage-biased Josephson junction drives the production of photons in a microwave resonator.
\end{abstract}

\vspace{2pc}

\section{Introduction}
\label{sec:Introduction}

The quantum micromaser consists of an optical cavity through which a stream of excited atoms pass and is a textbook example of a driven open quantum system\,\cite{Scully1997,Haroche2006,Englert2002}. Described by an elegant theoretical model\,\cite{Englert2002,Filipowicz1986}, the realization of the micromaser in experiment marked an important milestone for the development of quantum optics\,\cite{Walther2006}. The highly nonlinear nature of the micromaser leads to a rich range of behaviours including dynamical transitions, bistability and the existence of trapping states, where the system becomes restricted to an extremely small state space leading in principal to the stabilisation of strongly nonclassical states.

 Impressive progress over recent years in areas such as superconducting circuits\,\cite{scc} and optomechanics\,\cite{om} has opened new horizons in terms of the types of quantum dynamics that can be realised and the quantities that can be measured. However, the micromaser remains an important source of inspiration for theoretical and experimental work on nonlinear quantum systems. Indeed in the last few years, several different analogues of the micromaser have been proposed in the form of microwave circuits\,\cite{armour,you,marthaler} or optomechanical\,\cite{nation} devices, some of which have been realised\,\cite{astafiev,petta}. Such analogies build on a more general idea of exploiting Josephson systems as alternative 'artificial' versions of quantum optics\,\cite{Blais2004,Hauss2008,scc}.

\begin{figure}
\centering
\includegraphics[width=0.65\textwidth]{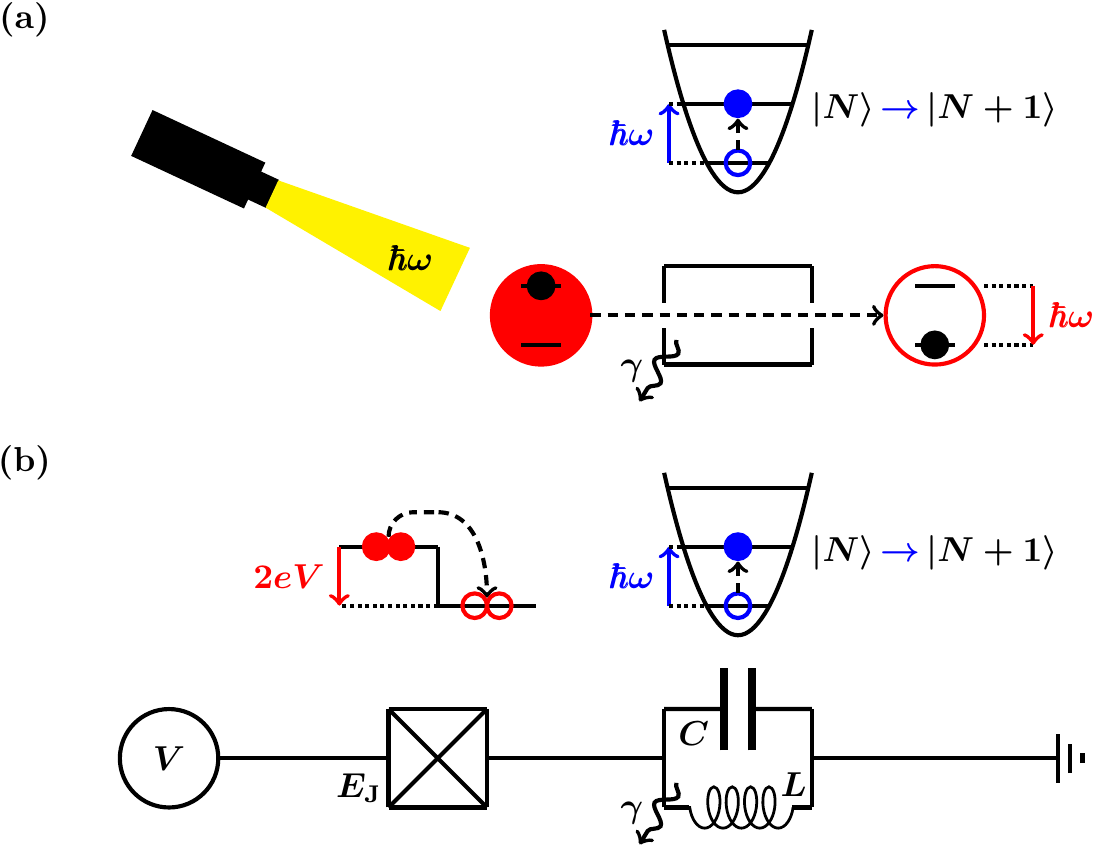}
\caption{Schematic set-up of (a) micromaser and (b) Josephson-photonics experiments. In the micromaser set-up, a microwave cavity is pumped by a continuous flow of excited atoms which interact strongly with one of the cavity modes. In Josephson-photonics systems, a Cooper-pair tunnelling across a dc-voltage--biased Josephson junction provides the energy to create a photon in a series-connected microwave resonator (or cavity). These creation processes are balanced by subsequent leakage of photons from the resonators after a lifetime $1/\gamma$ so that these highly nonlinear systems reach a steady state, offering strongly nonclassical states of light.}
\label{fig:Fig_1}
\end{figure}

 It is also fruitful to compare and contrast the micromaser with some of the novel paradigms for nonlinear open quantum systems offered by engineered quantum devices.
 In this article we focus on the behaviour of a Josephson-photonics device in which a flow of Cooper pairs across a voltage-biased Josephson junction (JJ) pumps a microwave cavity which is connected in series with the junction. Such systems have been explored in a range of recent experiments\,\cite{hofheinz,chen:2014,cassidy,westig17,amplifier,antibunched} and theoretical studies\,\cite{paduraiu:12,leppa:13,Armour2013,Gramich2013,clerk,switching,leppa18}. Although there is no direct mapping between the micromaser and Josephson-photonics devices the latter display many characteristics, such as dynamical transitions and trapping which are reminiscent of behaviour discussed before in the context of the micromaser. Hence, whilst they are perhaps not the closest of relatives the two types of system are at least cousins. As we seek to show below, comparing the similarities and differences between Josepshon-photonics devices and the long-studied, well-understood behaviour of the micromaser does not just deepen our understanding of the former, it also helps to point out new directions for future research.

 The two types of system we compare are illustrated schematically in Fig.\ \ref{fig:Fig_1}. In both cases a damped cavity mode is pumped by a flow of particles, two-level atoms in the case of the micromaser and Cooper pairs in Josephson-photonics set-ups. Even at this level, the apparent similarity between the two systems is off-set by an important difference: whilst the flow of charges through the Josephson circuit is determined by the interaction with the cavity, the centre of mass motion of the atoms in the micromaser is independent of the corresponding cavity dynamics. In the following we introduce the effective descriptions of the cavity mode dynamics in each case which then allows us to compare their properties in detail.

The rest of this paper is organised as follows. In section~\ref{sec:Model} we introduce and compare the theoretical model used for micromasers and Josephson photonics. We examine the impact of nonlinearities on the dynamics in Sec.~\ref{sec:Nonlinear}. Then we contrast the trapping states produced in the two systems in Sec.~\ref{sec:Non-classical}, and in Sec.~\ref{subsec:coherent} we discuss the ways in which cavity linewidths are accounted for in the two systems. We discuss the different ways in which information can be obtained about the state of the cavity in Sec.~\ref{sec:measure}.  Finally, Sec.~\ref{sec:Discussion} contains a short discussion and perspectives for future research.

\section{Model systems}
\label{sec:Model}

The basic interaction between the atoms and the cavity field in the micromaser is described by the Jaynes-Cummings model\,\cite{Filipowicz1986}, but a remarkably simple description of the system which involves the field mode alone can be obtained by coarse-graining over time\,\cite{Filipowicz1986,Englert2002}. Within the interaction picture, the master equation for the cavity density operator takes the form
\begin{equation}
\dot{\rho}={\mathcal{L}}_{a}[\rho]+{\mathcal{L}}_{d}[\rho] \label{eq:master1}
\end{equation}
where the first term describes the atom-field interactions,
\begin{equation}
\hspace*{-0.5cm}
{\mathcal{L}}_{a}[\rho]=N\left[\cos(\phi\sqrt{aa^{\dagger}})\rho\cos(\phi\sqrt{aa^{\dagger}})+\frac{a^{\dagger}\sin(\phi\sqrt{aa^{\dagger}})}{\sqrt{aa^{\dagger}}}\rho\frac{\sin(\phi\sqrt{aa^{\dagger}})a}{\sqrt{aa^{\dagger}}}-\rho\right]\\
\end{equation}
with $a$ the cavity mode lowering operator, $\phi$ the Rabi angle which is a product of the light-matter coupling and the time taken by the atoms to traverse the cavity\,\cite{Englert2002}, and $N$ the average flow rate of the atoms.
The second term in (\ref{eq:master1}) describes the usual dissipative interaction between a cavity mode and its surroundings, assumed to be at zero-temperature for simplicity,
\begin{equation}
{\mathcal{L}}_{d}[\rho]=\frac{\gamma}{2}\left[2a\rho a^{\dagger}-a^{\dagger}a\rho-\rho a^{\dagger}a\right], \label{eq:mmme}
\end{equation}
where $\gamma$ is the damping rate. Although the original atom-field interaction is coherent, the statistics of the atoms' arrivals at the cavity is Poissonian and when they are traced out the resulting master equation is fully incoherent.

For the simplest Josephson-photonics set-up, a cavity in series with a JJ biased at a voltage matching the fundamental mode frequency of the former [see Fig.\ \ref{fig:Fig_1}(b)], a simple effective description can also be derived in which only the cavity mode features. In a frame rotating at the cavity frequency, the resulting master equation is
\begin{equation}
\dot{\rho}=-\frac{i}{\hbar}[H_{\mathrm{JP}},\rho]+{\mathcal{L}}_{d}[\rho] \label{eq:master2}.
\end{equation}
In this case cavity dissipation is combined with a coherent interaction, which provides an approximate description of the effect of coherent Cooper pair tunnelling through the junction on the cavity mode\,\cite{Armour2013,Gramich2013}. The Hamiltonian takes the form
\begin{equation}
H_{\mathrm{JP}}=i\frac{E^{*}_{\rm J}}{2}:\left[a^{\dagger}-a\right]
\frac{{J}_1(2\Delta_0\sqrt{a^{\dagger}a})}{(a^{\dagger}a)^{1/2}}:, \label{eq:hrwa}
\end{equation}
where $E^{*}_{\rm J}={E}_{\rm J}{\rm{e}^{-\Delta_0^2/2}}$ is the Josephson energy of the function, renormalised by quantum fluctuations, and colons imply normal ordering.
The parameter $\Delta^{2}_0=4\pi Z_{\mathrm{LC}}/R_{\mathrm{K}}$ containing the impedance $Z_{\mathrm{LC}}$
of the superconducting resonator and the von Klitzing constant $R_{\mathrm{K}}=h/e^{2}$ represents the analogue to the fine structure constant of quantum electrodynamics, $\alpha=Z_{0}/(2R_{\mathrm{K}})$ with the vacuum impedance $Z_{0}$. It provides here a measure of the zero-point fluctuations of the oscillator.
As the effective fine structure constant of the model depends on the impedance $Z_{\mathrm{LC}} = \sqrt{L/C} $ [see Fig.~\ref{fig:Fig_1}(b)], and thus on circuit parameters, designing dedicated large-impedance devices allows one to reach a regime of strong-coupling quantum electrodynamics (with $\Delta^{2}_0\sim 1$ achieved in\,\cite{antibunched}).
In the limit of small $E_{\mathrm{J}}$ and $\Delta_0$, equation (\ref{eq:hrwa}) can be approximated as $H_{\mathrm{JP}}\simeq i(E^{*}_{\rm J}\Delta_0/2)(a^{\dagger}-a)$. Together with (\ref{eq:master2}), this describes an oscillator which is damped and driven linearly on resonance, leading to a steady state which is simply a coherent state\,\cite{Armour2013}.

\begin{figure}
\centering
\includegraphics[width=1.0\textwidth]{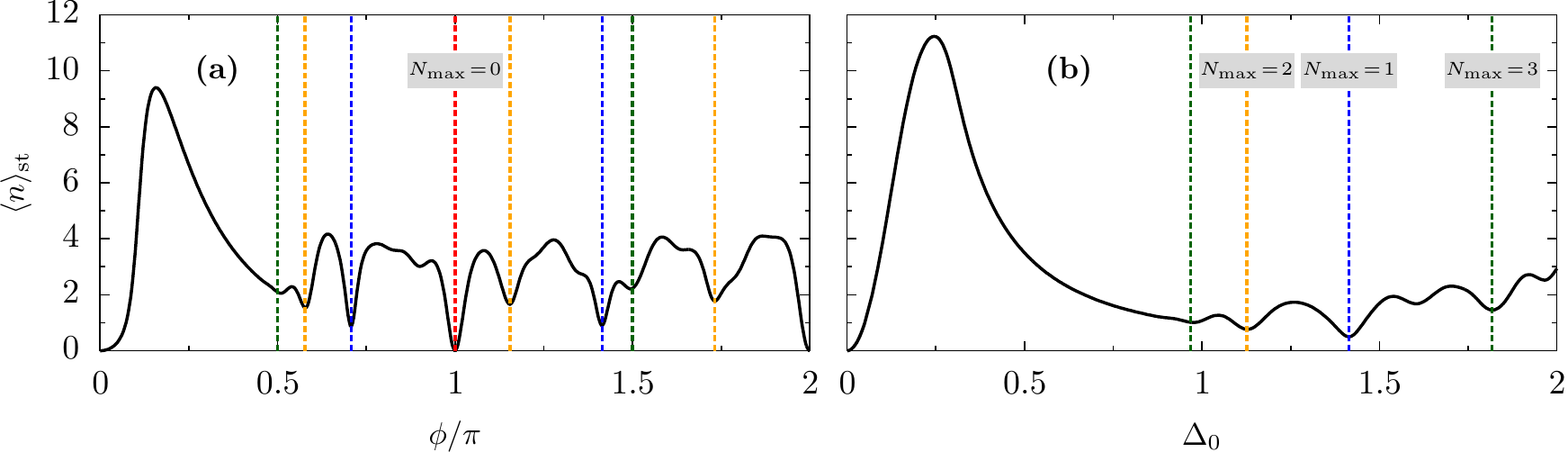}
\caption{Steady-state occupation of the (a) micromaser and (b) Josephson-photonics system as a function of the Rabi angle $\phi$ with pump parameter $N/\gamma=10$ and the strength of the zero-point fluctuations $\Delta_{0}$ with pump parameter $E^{*}_{\rm J}/(\hbar\gamma)=20$, respectively. Minima indicated by coloured vertical lines reveal the vanishing of transition matrix elements and thus the occurrence of few-level systems, where the number of maximal photons in the resonator is restricted to $0$ (red), $1$ (blue), $2$ (yellow), and $3$ (green).}
\label{fig:Fig_2}
\end{figure}

In both systems there are pairs of parameters that play complementary roles. One parameter, $\phi$ or $\Delta_0$ describes the strength of the underlying nonlinearity whilst another, $N/\gamma$ or $E_{\mathrm{J}}^*/\hbar\gamma$, acts like a pump rate. When the effective interaction strength reaches unity nonlinear effects become important for occupation numbers of order unity, but at much lower effective interaction strengths nonlinear effects come into play when the pumping parameter is increased enough to generate sufficiently large photon populations. As can be seen in Fig.\ \ref{fig:Fig_2}, whilst increasing $\phi$ ($\Delta_0$) from zero naturally leads to an increase in steady-state occupation numbers, larger effective interaction strengths eventually lead to a systematic lowering of the occupation numbers.

\section{Nonlinear dynamics}
\label{sec:Nonlinear}

A striking feature of both the micromaser and Josephson-photonics devices is the emergence of transcendental functions of the number operator in the effective master equation descriptions. This highly nonlinear dependence leads naturally to a very rich nonlinear dynamics which emerges clearly at the mean-field or semiclassical level\,\cite{Haroche2006,Filipowicz1986,Armour2013}. For a Josephson-photonics device in the semiclassical limit, where the system is described by amplitude $A$ and phase $\phi$ given by $\langle a\rangle=Ae^{i\phi}$, one finds the Hamiltonian
\begin{equation}
H_{\mathrm{sc}}=E^{*}_{\mathrm{J}}\sin{(\phi)}J_{1}(2\Delta_{0}A),
\end{equation}
which is nothing but a nonlinearly driven harmonic oscillator on resonance\,\cite{switching}. The corresponding equations of motion (including damping), take the form
\begin{equation}
\dot{\phi}=-\frac{E_{\mathrm{J}}^{*}\Delta_{0}}{A}\sin{(\phi)}J'_{1}(2\Delta_{0}A),\quad \dot{A}=-\frac{\gamma}{2}A+\frac{E^{*}_{\mathrm{J}}}{2A}\cos{(\phi)}J_{1}(2\Delta_{0}A).
\end{equation}
From the phase equation, we see that either the phase or the amplitude of each of the fixed points must be `locked' in the sense that it does not vary with $E_{\mathrm{J}}$. A more detailed analysis shows that as $E_{\mathrm{J}}$ is increased the phase of the steady-state (i.e.\ the stable fixed point) is locked initially, but at larger $E_{\mathrm{J}}$ a bifurcation arises leading to an amplitude-locked solution\,\cite{switching}. Such dynamical transitions, which look more like traditional phase transitions with sharp features rather than smooth crossovers in the semiclassical limit where photon numbers are large, were also found for micromasers\,\cite{Filipowicz1986}.

The strong nonlinearity and relatively weak noise in both the micromaser and Josephson-photonics devices means that they display signatures of nonclassicality even in regimes where the photon numbers are large. In particular, both exhibit number-squeezing over a wide range of parameters such that the width of the number state distribution is less than that of a coherent state with corresponding average occupation: $F_n=(\langle n^2\rangle-\langle n\rangle^2)/\langle n\rangle< 1$, $n=a^{\dagger}a$. Such squeezing does not happen across the whole parameter space, with strong peaks in $F_{n}$ signifying the onset of dynamical transitions in the micromaser\,\cite{Filipowicz1986}. Interestingly, the pattern of dynamical transitions in JJ-resonator systems can be quite different, leading to much weaker signatures in the fluctuations in the number-state distribution\,\cite{Armour2013}.

One of the most attractive aspects of the micromaser model (\ref{eq:master1}) is that it allows for a full analytic solution for the steady-state elements of the density matrix. This is because the micromaser master equation leads to dynamics in which the diagonal and off-diagonal elements in the Fock-state basis decouple. This means that the off-diagonal elements vanish in the steady state and the application of detailed balance\,\cite{Scully1997,Englert2002} leads to the following expression for the diagonal ones
\begin{equation}
\langle n|\rho|n\rangle=n_0\prod_{m=1}^{n}\frac{N}{\gamma}\frac{\sin^2(\phi\sqrt{m})}{m},
\end{equation}
with $n_0$ a normalization constant.
Unfortunately no such compact analytic expression has (yet) been obtained for the steady state for Josephson-photonics devices, largely because of the complications caused by the coherent dynamics in the system. Consequently, we have to rely here on standard numerical methods to find the steady state of the quantum master equation [Eq.~(\ref{eq:master2}) with the rotating-wave approximation Hamiltonian in Eq.~(\ref{eq:hrwa}) and dissipator in Eq.~(\ref{eq:mmme})], allowing us to study the corresponding steady-state photon occupation numbers and Wigner densities.

\section{Trapping states}
\label{sec:Non-classical}

The nonlinearities in both the micromaser and JJ-cavity system can also be exploited to stabilise states in which the cavity becomes trapped in the sense that its dynamics cannot access Fock states above a specific value. Despite this apparent similarity, the properties of the resulting trapped states differ in important ways. In particular, the dissipative form of the master equation for the micromaser actually makes it significantly easier to stabilise Fock-like states via the trapping mechanism.

In the case of the micromaser, the uncoupling of the diagonal and off-diagonal elements of the density operator in the Fock state basis leads to a simple dynamics consisting of upward and downward transitions between the  diagonal elements (populations).
In the zero temperature limit, the cavity is only ever damped by its surroundings, leading to transitions which are always downwards. The interaction between the cavity and the atoms, described by $\mathcal{L}_a$ in (\ref{eq:master1}), generates upward transitions, but the corresponding rates can vanish for special values of $\phi$ at which $\sin(\phi\sqrt{N_{\rm max}+1})=0$ with $N_{\rm max}=0,1,2,...$. In such cases the micromaser can relax from higher to lower number states, but it can never make an upward transition between the $N_{\rm max}$ and $N_{\rm max}+1$ states, hence the steady-state probability of finding the system in a number state $n>N_{\rm max}$ vanishes.

Something very similar happens in the case of Josephson photonics\,\cite{Gramich2013,clerk}. Here the Hamiltonian generated evolution is coherent and
the matrix elements between different number states of the cavity take the form
\begin{equation}
\langle n|H_{\mathrm{JP}}|n+1\rangle=-i\Delta_0\frac{{E^{*}_{\rm J}}}{2\sqrt{n+1}}L_n^1(\Delta_0^2), \label{eq:matrixelements}
\end{equation}
whilst elements beyond the first off-diagonal are all zero.
  Provided the zero-temperature limit is taken, the effect of the cavity's surroundings is again limited to producing downward transitions in the number state basis. The system becomes trapped when the parameter $\Delta_0$ is tuned to a zero of the associated Laguerre polynomial: $L_{N_{\rm{max}}}^1(\Delta_0^2)=0$ implies that the steady state of the system will not involve number states with $n>N_{\rm{max}}$.

Trapping within a restricted Fock state basis is therefore a feature of both systems. Fig.\ \ref{fig:Fig_2} illustrates values of the corresponding device parameters $\phi$ at $\Delta_0$ which trapping occurs below the Fock state $N_{\rm max}=0,1,2,3$. Trapping appears as a localised dip in the steady state occupation number $\langle n\rangle_{\rm st}$ at the appropriate parameter value. Interestingly, in contrast to the micromaser, the JJ-cavity system cannot be trapped in the ground state because $L_0^1(x)=1$. The transition matrix element $\langle n|H_{\mathrm{JP}}|n+1\rangle\propto\langle n|\exp{[i\Delta_{0}(a^{\dagger}+a)]}|n+1\rangle\propto L_n^1(\Delta_0^2)$ corresponds to the overlap of the shifted harmonic-oscillator eigenstates of the involved levels. The mathematical observation $L_0^1(x)=1$ thus reflects the physical fact that the overlap between the harmonic-oscillator wave functions of the ground state and first excited state does not vanish for any finite shift.

\begin{figure}
\centering
\includegraphics[width=0.4\textwidth]{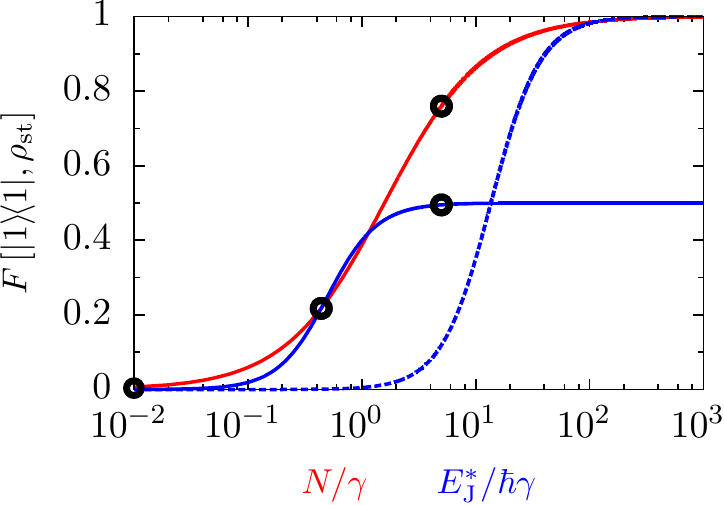}
\caption{Fidelity $F$ of Fock state $\ket{1}\!\!\bra{1}$ and steady state $\rho_{\mathrm{st}}$ of the micromaser (red) and the standard single-cavity Josephson system (blue, solid) as a function of pump parameters $N/\gamma$ and $E^{*}_{\rm J}/(\hbar\gamma)$, respectively, in the two-level situation ($\phi=\pi/\sqrt{2}$, $\Delta_{0}=\sqrt{2}$). In contrast to the single-cavity Josephson-photonics system, the micromaser reaches population inversion, $\langle n\rangle_{\mathrm{st}}>0.5$, for sufficiently strong driving so that Fock states $\ket{1}\!\!\bra{1}$ with high fidelity can be created. This can also be realized within the Josephson-photonics platform on the basis of a two-cavity scheme (cf. Ref.\,\cite{clerk}) when using an auxiliary cavity with photon leakage rate $\gamma_{\mathrm{aux}}=100\gamma$ (blue, dashed). The Wigner densities of the states associated with the parameters indicated by black circles are pictured in Fig.~\ref{fig:Fig_4}.}
\label{fig:Fig_3}
\end{figure}

\begin{figure}
\centering
\includegraphics[width=1.0\textwidth]{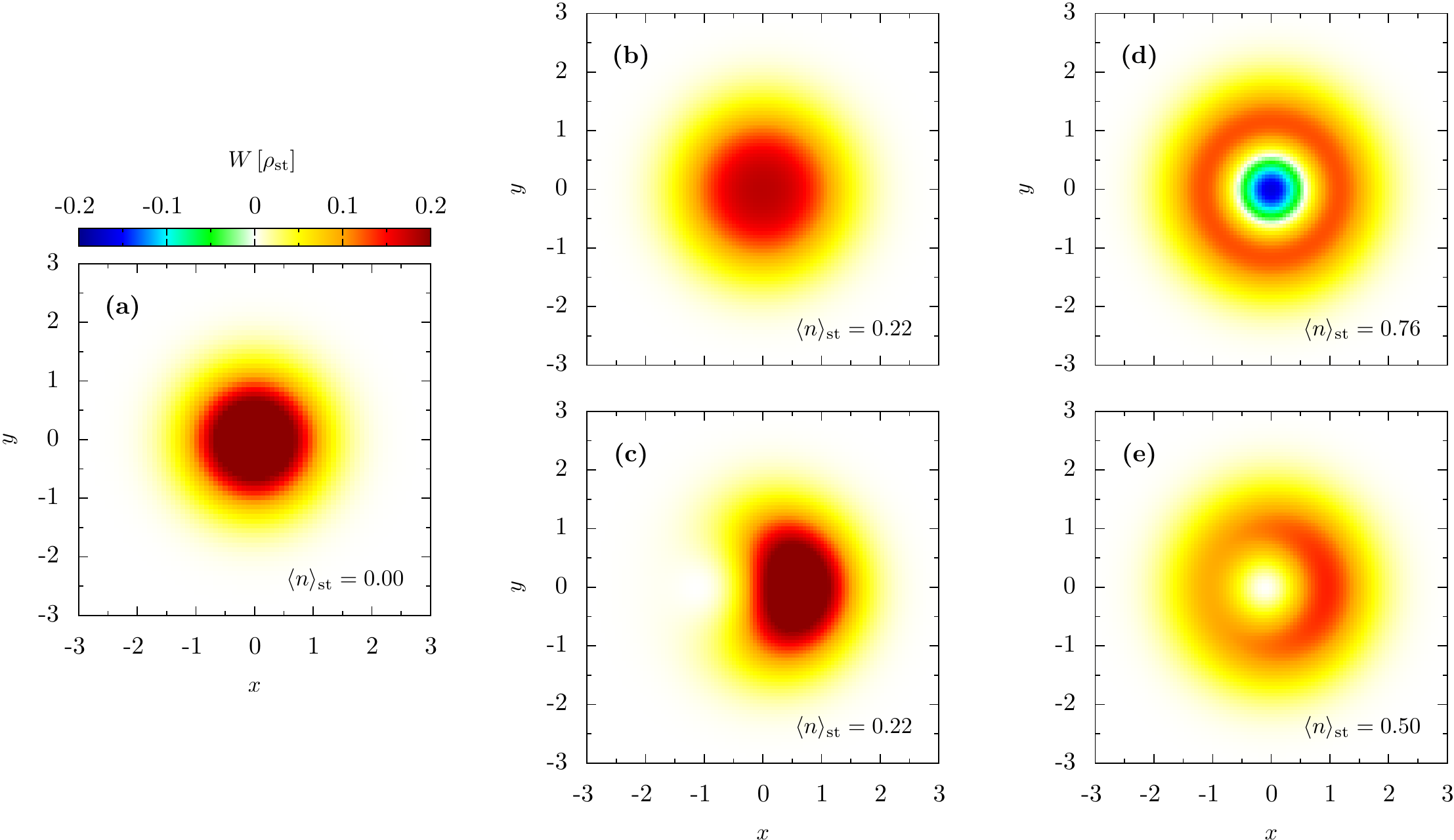}
\caption{Wigner functions for the steady state $\rho_{\mathrm{st}}$ of the micromaser system [(a), (b), and (d)] and the Josephson-photonics system [(a), (c), and (e)] for different steady-state occupation numbers (cf. black circles in Fig.~\ref{fig:Fig_3}).}
\label{fig:Fig_4}
\end{figure}

The most significant difference in the way in which trapping occurs between the two systems stems from the fact that trapping arises through the vanishing of matrix elements in the Hamiltonian for the JJ-cavity system and of incoherent transition rates in the case of the micromaser. As Fig.\ \ref{fig:Fig_3} illustrates for the case where $N_{\rm max}=1$, in both cases the system can be driven closer towards a pure Fock state by increasing the corresponding drive parameter ($N/\gamma$ or $E_{\mathrm{J}}^*/\hbar\gamma$). However, whilst one can in principal get arbitrarily close to the Fock state $|1\rangle$ for the micromaser, the coherent interaction in the JJ-cavity system sets a fundamental limit of 0.5 on the fidelity. The effect that this has on the actual states that are produced is shown in Fig.\ \ref{fig:Fig_4}. In particular, the trapping states produced in the JJ-cavity system are too far from being a Fock state for the corresponding Wigner functions to display negative regions\,\cite{clerk}.
Nevertheless, the fact that trapping with $N_{\rm max}=1$ can be achieved means that both systems could function as sources of single photons\,\cite{mmsps,antibunched}.
As pointed out in Ref.\,\cite{clerk}, an almost perfect Fock state can be produced (see Fig.\ \ref{fig:Fig_4}) by means of a second cavity which is much more strongly damped and acts to degrade unwanted coherent oscillations.

\section{Cavity linewidth}
\label{subsec:coherent}

The effective dissipative master equation for the micromaser cavity leads to laser-like dynamics in which the system undergoes phase diffusion\,\cite{Scully1997} (i.e.\ a finite linewidth), so that it always eventually evolves into a state which has a circularly symmetric Wigner function. In contrast, in Josephson photonics the action of the Hamiltonian in the master equation leads to a dynamics which is apparently that of nonlinear resonance: the system has a preferred phase and the linewidth vanishes.

The spectral properties of the micromaser were investigated by Wolfgang Schleich and co-workers during the 1990s using a variety of different methods (see, e.g.\,\cite{schleich1,schleich2}). The complex nonlinear dynamics of the system are reflected in the behaviour of the phase diffusion which shows features quite distinct from those seen in the standard laser\,\cite{schleich2}.

In a cavity-JJ system the preferred phase is ultimately set by the voltage source which acts on the cavity like a time dependent drive. The effective Hamiltonian (\ref{eq:hrwa}) emerges within a rotating frame after a rotating wave approximation is made and the original Hamiltonian has an explicit time dependence. However, experiments on Josephson cavity systems do not reflect this picture\,\cite{hofheinz,chen:2014,cassidy} and in fact reveal behaviour which is much more like a laser in that the steady states have no preferred phase.

The simplicity of the resonator-only effective model, (\ref{eq:master2}), is achieved by suppressing an aspect of the physics which in this context proves important: in any real circuit the effective voltage bias experienced by the resonator will be subject to low frequency fluctuations\,\cite{Gramich2013,clerk,wang17}. This defect of the model can be remedied by extending the Hilbert space to include the charges transported across the junction which experience an effective dephasing due to voltage fluctuations\,\cite{Gramich2013}. Within a resonator-only picture voltage fluctuations can be accounted for as an additional classical noise source in a Langevin description\,\cite{wang17} or through additional terms in the master equation describing photon-number dephasing\,\cite{clerk}. In any case the linewidth of the system is set by the low frequency voltage fluctuations which (at least so far) is regarded effectively as a parameter that simply needs to be added to the theory. In the case of the micromaser, the master equation explicitly accounts for a quite different classical stochastic effect: random arrival times of the atoms at the cavity.

A proper description of the spectral broadening due to voltage noise is crucial for describing the emission spectrum of the resonator which can contain nontrivial information about the system dynamics. As an example, we show in Fig.~\ref{fig:Fig_5}(a) calculations of the power spectral density, developing a characteristic Mollow-triplet structure, as the strongly-nonlinear system ($\Delta_0=1$ as reached in recent experiments\,\cite{antibunched}) is driven strongly. For this case, voltage fluctuations can be treated as quasi-static, so that a classical average over a Gaussian distribution of voltages (with parameters extracted from the linear-driving spectrum) can model the expected spectrum.

Whilst low-frequency voltage fluctuations play a crucial role in Josephson photonics in destroying preferred phase
and broadening emission spectra, their influence on the energetics of the system is much less significant and in many cases they can be neglected\,\cite{Gramich2013,clerk,switching}. In regimes where the resonator occupation remains small, the voltage fluctuations have been shown explicitly to lead to a small correction\,\cite{Gramich2013}, though they are expected to become more important in the semiclassical limit where photon numbers become large\,\cite{wang17}.

\section{Probing the cavity state}
\label{sec:measure}

\begin{figure}
\centering
\includegraphics[width=1.0\textwidth]{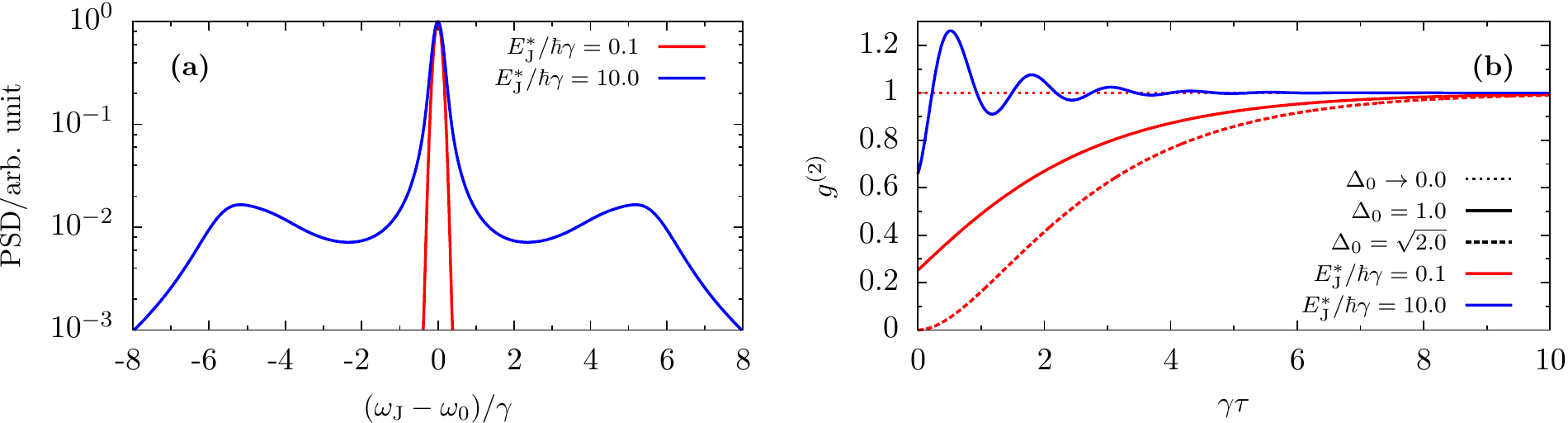}
\caption{(a) Power spectral density (PSD) in arbitrary units (normalized to the value of the main peak at $\omega_{\rm J}=\omega_{0}$) for $\Delta_{0}=1$ and two different values of $E^{*}_{\rm J}/(\hbar\gamma)$. Here, we have included voltage fluctuations into our description by averaging over a Gaussian distribution of static noise with standard deviation $\sigma^{2}=\gamma/10$. (b) Second-order correlation function $g^{(2)}$ for different values of the strength of the zero-point fluctuations $\Delta_{0}$ and driving strength $E^{*}_{\rm J}/(\hbar\gamma)$.
The experimentally realized strong anti-bunching for  $\Delta_{0}$ (cf.\,\cite{antibunched}) lies between the well-known results for the harmonic limit, $\Delta_0 \rightarrow 0$, and the two-level systems realized as  $\Delta_0 =\sqrt{2}$.
}
\label{fig:Fig_5}
\end{figure}

In the micromaser information about the dynamics of the system is obtained by measuring the state of atoms emerging from the cavity. Detailed descriptions of the statistics of the clicks\,\cite{Briegel1994,Englert2002} of an atom-state detector have been developed including full counting statistics and waiting time distributions. Nevertheless, despite their practical advantages, atom-state measurements do not provide simple answers about all of the cavity properties of interest\,\cite{Briegel1994}.

In Josephson photonics the practical situation is very different. Coupling of the resonator to a transmission line is straightforward, providing direct access to photonic properties through the radiation that leaks out of the cavity. Such leakage damps the cavity and indeed is typically the dominant mechanism (i.e.\ it contributes essentially all of the rate $\gamma$). Measurements carried out so far include\,\cite{hofheinz,chen:2014,cassidy,antibunched} the average power and power spectral density,
 and second-order coherence, $g^{(2)}(0)=\langle n(n-1)\rangle/\langle n\rangle^2$. Indeed, predictions of number squeezing in the cavity\,\cite{Gramich2013,Armour2013} have recently been confirmed by detecting values\,\cite{antibunched} of $g^{(2)}(0)<1$ (which implies $F_n<1$). In Fig.~\ref{fig:Fig_5}(b) we show anti-bunching and Rabi-type oscillations in $g^{(2)}(\tau)$ for various parameter regimes, including the scenario discussed above via its first-order correlations, see Fig.~\ref{fig:Fig_5}(a). The reduction to a two-level system for $\Delta_0 = \sqrt{2}$ results in perfect anti-bunching, but drastic signatures of a nearly-trapped system [cf. Sec. 4 and Fig.~\ref{fig:Fig_2}(b)] remain clearly visible well below this value.

The fact that essentially all of the radiation leaking out of superconducting microwave cavities can be captured and detected suggests that the full counting statistics of the photons should in principal be measurable\,\cite{paduraiu:12}. In this case, the earlier work on the statistics of detector clicks\,\cite{Briegel1994} provides clear inspiration and this is an area in which further theoretical work is to be expected in the future.

In Josephson-photonics set-ups the current flowing through the Josephson junction can also be readily measured, providing a second channel of information about the dynamics of the system. Unlike the centre-of-mass motion of the atoms through the micromaser cavity, net transport of Cooper pairs through the JJ implies photon generation. Experiments so far have demonstrated the expected relationship between the average current and the rate at which photons are produced\,\cite{hofheinz}. The corresponding fluctuations contain further dynamical information which is expected to be directly related to the fluctuations in the photons leaking out of the cavity\,\cite{switching}.

\section{Discussion}
\label{sec:Discussion}

In our comparison of the micromaser and Josephson-photonics devices we have seen that many of the most interesting features in the latter have parallels that were already discussed and understood in the context of the former. In particular, both systems display interesting dynamical transitions, give rise to number squeezing, display trapping states  and are subject to phase diffusion.

Despite these similarities there are also essential differences. Most fundamental is the different nature of the objects described by the respective theories. While the atoms and cavity photons in the micromaser are truly microscopic objects, the degrees of freedom of Josephson photonics are collective entities emerging as macroscopic quantum variables of a lumped-circuit description of the full microscopic many-body Hamiltonian of the actual condensed-matter device. The fact, that in Josephson photonics we are dealing with an effective theory, is directly linked to the possibility to change fundamental parameters of the theory, such as the effective fine structure constant, by circuit design.

Realizing a condensed-matter cousin of the micromaser, thus, not only offers new technological aspects, such as the easy access to the photonic properties of on-chip microwave cavities already discussed above. Much more significantly,  the fact that both the impedance of superconducting resonators and the Josephson energy of junctions can be varied within very wide ranges in experiments\,\cite{chen:2014,cassidy,antibunched} means that the strength of the nonlinear interactions that are generated can be made larger than the level spacing of the cavity. This should allow access to novel regimes of ultrastrong coupling for which the master equation (\ref{eq:master2}) will surely prove inadequate. In contrast, fundamental atomic physics dictates that the strength of the underlying light-matter coupling in the micromaser is very weak. Although the {\emph{effective} interaction can become very strong (by tuning the time spent by an atom traversing the cavity so that the probability of transferring energy to an unoccupied cavity can reach unity), the weakness of the bare interaction means that the rotating wave approximation that goes into the Jaynes-Cummings description of the atom-cavity dynamics remains valid and a simple description of the cavity losses remains appropriate. In Josephson photonics, the rotating wave description that underlies the effective description (\ref{eq:master2}) is expected to fail for ultrastrong coupling.

\ack
This work was supported by the Deutsche Forschungsgemeinschaft (DFG) through Grant No. AN336/11-1 and by the Center for Integrated Quantum Science and Technology (\smash{$\mathrm{IQ}^{\mathrm{ST}}$}). ADA was supported by a Leverhulme Trust Research Project Grant (RBG-2018-213) and SD acknowledges financial support from the Carl Zeiss Foundation.

\end{document}